\documentclass[useAMS,usenatbib]{mn2e}   

\usepackage{amsmath}

\usepackage{natbib, txfonts, latexsym, lscape}
\usepackage{graphicx}

\newcommand{\msun}{$h^{-1}{\rm M}_{\odot}$}

\newcommand{\per}{$^{-1}$}
\newcommand{\hinv}{$h^{-1}$}

\title[Disentangling galaxy environment and host halo mass]{Disentangling galaxy environment and host halo mass}
\author[M.R. Haas, J. Schaye and A. Jeeson-Daniel]{Marcel R. Haas$^{1,2}$\thanks{E-mail: mhaas@stsci.edu (MRH)}, Joop Schaye$^{1}$ and Akila Jeeson-Daniel$^{1,3}$ \\
$^{1}$Leiden Observatory, Leiden University, P.O. Box 9513, NL-2300 RA, Leiden, The Netherlands \\
$^{2}$Space Telescope Science Institute, 3700 San Martin Drive, Baltimore, MD 21218, USA \\
$^3$Max-Planck-Institut f\"ur Astrophysik, Karl-Schwarzschild-Str. 1, D-85748, Garching, Germany}

\begin{document}

\date{Accepted Not yet. Received not yet; in original form \today}

\pagerange{\pageref{firstpage}--\pageref{lastpage}} \pubyear{2011}

\maketitle

\label{firstpage}


\begin{abstract}
The properties of both observed galaxies and dark matter haloes in simulations depend on their environment. The term ``environment'' has, however, been used to describe a wide variety of measures that may or may not correlate with each other. Popular measures of environment include, for example, the distance to the $N^{\rm th}$ nearest neighbour, the number density of objects within some distance, or, for the case of galaxies only, the mass of the host dark matter halo. Here we use results from the Millennium simulation and a semi-analytic model for galaxy formation to quantify the relations between different measures of environment and halo mass. We show that the environmental parameters used in the observational literature are in effect measures of halo mass, even if they are measured for a fixed stellar mass. The strongest correlation between environmental density and halo mass arises when the number of objects is counted out to a distance of 1.5 -- 2 times the virial radius of the host halo and when the galaxies/haloes are required to be relatively bright/massive. For observational studies this virial radius is not easily determined, but the number of neighbours out to 1 -- 2 \hinv Mpc gives a similarly strong correlation with halo mass. For the distance to the $N^{\rm th}$ nearest neighbour the (anti-)correlation with halo mass is nearly as strong provided $N\ge 2$. We demonstrate that this environmental parameter becomes insensitive to halo mass if it is constructed from dimensionless quantities. This can be achieved by scaling the minimum luminosity/mass of neighbours to that of the object that the environment is determined for and by dividing the distance to a length scale associated with either the neighbour or the galaxy under consideration. We show how such a halo mass independent environmental parameter can be defined for both observational and numerical studies. The results presented here will help future studies to disentangle the effects of halo mass and external environment on the properties of galaxies and dark matter haloes.
\end{abstract}


\begin{keywords}
galaxies: haloes -- galaxies: statistics -- galaxies: fundamental parameters -- galaxies: evolution -- galaxies: general -- methods: statistical
\end{keywords}


\section{Introduction}
The formation and evolution of galaxies depends on both internal and external processes (`nature vs. nurture'). Among the internal processes are radiative cooling, the formation of and feedback from stars, and accretion of gas onto and feedback from super-massive black holes. It is generally assumed that halo mass is the fundamental parameter that drives the internal processes for isolated galaxies. External processes are important because galaxies do not live alone in the Universe. Galaxy interactions can induce gravitational torques that can significantly alter the angular momentum structure of the matter in galaxies. This can for example lead to a starburst or to more rapid accretion onto the central black hole, which may trigger a quasar phase. Smaller galaxies may accrete onto the halo of a more massive galaxy. As a galaxy moves through the gaseous halo of a more massive galaxy, it may lose gas due to ram pressure forces. Winds and radiation from nearby neighbours may also affect the evolution of a galaxy. It is still an open question to what extent the properties of galaxies are determined by internal and external processes. 

Even if halo mass were the only driver of galaxy evolution, galaxy properties would still be correlated with their environment. Because peaks in the initial Gaussian density field cluster together, more massive galaxies will live close to each other (`galaxy bias'). A correlation between surrounding galaxy density and internal galaxy properties therefore does not necessarily imply a causal relation between the two. %

Early, analytic models predicted that the clustering of haloes depends
only on their mass \citep{kaiser84, colekaiser89, mowhite96}.
\citet{lemsonkauffmann99} found in semi-analytic models of galaxy
formation that, to first order, the only property of a dark matter halo that correlates
with the (projected) number density of surrounding galaxies is halo mass. Other properties like  spin parameter,
formation time and concentration do not strongly depend on the
surrounding dark matter density. However, later papers have shown that
clustering also depends on properties like formation time, concentration, substructure content, spin and shape,
even for fixed mass \citep[e.g.][]{gao05, harker06, wechsler06, bett07,
gaowhite07, jing07, maccio07, wetzel07, angulo08,
faltenbacherwhite10}. All dependencies other than the one with halo
mass are, however, second-order effects.  
The formation time and the
halo merger rate are also found to depend only weakly on environment
\citep{gottloeber01, shethtormen04, fakhourima09, hahn09}. These findings are consistent with the
hydrodynamical simulations of \citet{crain09}, who found that all the
variations in the properties of simulated galaxies with environment can be accounted for by the dependence of the halo mass function on
environment.

For both observations and simulations it is difficult to disentangle the effects of halo mass from those of the external environment. The two are correlated (higher mass haloes live, on average, in denser environments) and finding an environmental parameter that does not correlate with halo mass is non-trivial. Of course, the mass of the dark matter halo hosting a galaxy is important for the evolution of that galaxy, so halo mass is as good an environmental parameter as any other. One would, however, like to be able to distinguish halo mass (the ``internal environment'') from the environment on large scales (the ``external environment''). It is not a priori clear whether the environmental parameters used in the literature measure halo mass, and if so, whether they measure \textit{only} halo mass, or whether they are also, or predominantly, sensitive to the external environment.

Observationally, halo mass is hard to determine. Group catalogues, abundance matching, clustering, and gravitational lensing all provide statistical measures of halo mass. Nonetheless, most observational data sets will have to do without dark matter halo mass and define environmental parameters based on the distribution of visible matter (usually stellar luminosity). 

Many observational studies have, nevertheless, investigated the effect of the environment on the physical properties of galaxies. In general, in higher density environments galaxies form their stars earlier and faster \citep[e.g.][]{lewis02, baldry04, balogh04, balogh04a, kauffmann04, thomas05, smith06} and galaxy morphologies become more (pressure support dominated) early type, as opposed to (rotation dominated) late type \citep[e.g.][]{dressler80, dressler97, wilman09}. From observations alone it is very hard to judge whether these trends are driven mostly by halo mass or whether other halo properties and/or large-scale environment play an important role. As in observations environment is usually contrasted with stellar mass (rather than halo mass), such observationally based distinctions between stellar mass and environment may tell us more about the stellar mass -- halo mass relation than about the difference between external environment and halo mass.

In simulations, halo mass (and other halo parameters) are readily available. From simulations much `cleaner' definitions of environment can be obtained, as the distance between objects is known in three dimensions, contrary to observations which can only provide precise separations perpendicular to the line of sight. Radial velocity differences give an indication of separations along the line of sight, but peculiar velocities complicate their interpretation.

Many different measures of environment have been used in the literature. Some are closely related by construction, while the relation between others is more obscure. In this paper we compare several popular indicators of environments. The aim is to investigate which indicators correlate strongly with each other and with halo mass and which ones do not. We measure environmental parameters using a semi-analytic model for galaxy formation \citep{deluciablaizot07} constructed on the merger tree of dark matter haloes formed in the Millennium Simulation \citep{millennium}, so that we also have halo masses available. We will present environmental parameters that measure halo mass, but are insensitive to external environment, along with environmental parameters that are insensitive to halo mass. These can be used for studies that aim to separate the effect of halo mass and external environment. We will show that most of the environmental indicators used in literature are in effect measures of halo mass. In the remainder of the paper we will use the term `environment' whenever we quantify distances to nearby galaxies, surrounding galaxy densities etc., but never when referring to halo mass, in order to clearly distinguish the two.

This paper is organised as follows. Section~\ref{sec:literature} gives a short overview of the literature on environmental parameters, both from observations and simulations. In Section~\ref{sec:sims} we investigate how often used environmental parameters correlate with host halo mass. The strength of the correlation with halo mass depends on the distance scale used in the environmental parameters, as we will show in Section~\ref{sec:distdep}. In Section~\ref{sec:massindep} we discuss how to construct an environmental parameter that is insensitive to halo mass. Finally, we conclude in Section~\ref{sec:conclusions}. In appendix~\ref{sec:mhalo_fits} we provide fitting functions for the host halo mass as a function of galaxy environmental parameters.


\section{Popular environmental parameters} \label{sec:literature}

The study of the effect of environment on the evolution of galaxies has undergone considerable progress through large galaxy surveys like the Sloan Digital Sky Survey \citep[SDSS;][]{stoughton02} and (z)COSMOS \citep{scoville07, lilly07}.
Many different definitions of environmental density exist. Observationally, the density around galaxies must usually be based on the distribution of the galaxies themselves, as the distribution of mass is very hard to measure. Two slightly different measures are used very often: the number of galaxies within a fixed distance and the distance to the $N^\textrm{th}$ nearest neighbour. Table~\ref{tab:literature} contains a short summary of the literature on the environmental dependence of galaxy properties, both from observations and from simulations. We will expand on these in this section in the next section we will study some of these in more detail using the galaxy catalogues from the Millennium database.

For the environmental parameters it is important, as we will show below, whether the masses of the other galaxies used to measure the environment have a fixed physical lower limit (or luminosity), or whether the minimum mass is a fixed fraction of the mass of the galaxy one wants to know the environment of. It also matters whether the distance out to which the environment is measured is fixed in absolute terms or whether it is fixed relative to some length scale related to the galaxy in question (e.g.\ the virial radius of its host halo). In Table~\ref{tab:literature} we indicate for each environmental parameter listed (described in the first column) out to what distance (or distance equivalent parameter) the environment is measured (second column), and whether the minimum mass/luminosity of the galaxies used for the environmental estimate is fixed in absolute terms or whether it is a fixed fraction of the mass/luminosity of the galaxy in question (if applicable, third column). The last column lists references to papers employing the parameter. From Table~\ref{tab:literature} it is clear that very few papers take minimum masses of neighbours and/or distances relative to properties of the galaxy's host halo.

\begin{table*}
\caption{Overview of environmental parameters that are frequently used in literature. They are grouped by the different ways of determining out to which distance the environment is measured either in observational or simulation studies. The first column specifies the environmental parameter, and the second and third column indicate out to what distance the environment is measured and whether the minimum mass/luminosity is fixed or scales with the galaxy in question. The fourth column specifies the references for the papers: 1: \citet{dressler80}, 2: \citet{whitmoregilmore91}, 3: \citet{gomez03},  4:\citet{postmangeller84}, 5: \citet{whitmore93}, 6: \citet{goto03}, 7: \citet{weinmann06}, 8: \citet{cooper05}, 9: \citet{cooper06}, 10: \citet{cooper08}, 11: \citet{balogh04}, 12: \citet{balogh04a}, 13: \citet{baldry06}, 14: \citet{bamford09}, 15: \citet{cassata07}, 16: \citet{ellison10},  17: \citet{kovac2010}, 18: \citet{pimbblet02}, 19:  \citet{lewis02}, 20: \citet{blanton05}, 21: \citet{blanton03}, 22: \citet{blanton03a}, 23: \citet{hogg03}, 24: \citet{blantonberlind07}, 25: \citet{berrier10}, 26: \citet{wilman10}, 27: \citet{hogg04}, 28:\citet{kauffmann04}, 29: \citet{parkchoi09}, 30: \citet{harker06}, 31: \citet{hestertasitsiomi10}, 32: \citet{maccio07}, 33: \citet{maulbetsch07}, 34: \citet{espinobriones07}, 35: \citet{ishiyama08}, 36: \citet{abbassheth05}, 37: \citet{crain09}, 38: \citet{lemsonkauffmann99}, 39: \citet{fakhourima09}, 40: \citet{hahn09}, 41: \citet{wang07}, 42: \citet{faltenbacher10}}.             
\label{tab:literature}      
\centering                          
\begin{tabular}{l l l l}        
\hline                
\hline
Parameter & Distance related parameter value & Minimum mass/luminosity & References \\
\hline
\hline
\textit{From observations} &  &  &  \\
\hline
(Projected) galaxy number density & Average of nearest 10 galaxies & $m_V < 16.5$ & 1, 2, 3 \\
                                                                &  & $M_V < -20.4$ & 3 \\
                                                   & Group average & $M_B < -17.5$                                     & 4 \\
\hline
Cluster/Group-centric radius & - &  $M_r < -20.5$ & 5, 6 \\
                             & - &  $M_V < -20.4$& 3 \\
                             & - &  $m_V < 16.5$ & 2 \\
   & Scaled to the virial radius & $r<17.77$ & 7 \\
\hline
Projected galaxy number density out                  & $N=$ 3, $\Delta v = $ 1000 km s$^{-1}$ & $R < 24.1$ & 8, 9, 10  \\
\,\,\,\,\, to the $N^\textrm{th}$ nearest neighbour  & $N=$ 4,5                                & $M_R < -20$ & 11 - 16  \\
\,\,\,\,\, with a maximum radial velocity            & $N=$ 4,5, $\Delta v = $ 1000 km s$^{-1}$ &  $M_r < -20$   &  13, 14 \\
\,\,\,\,\, difference $\Delta v$                     & $N=$ 4,5, $\Delta v = $ 1000 km s$^{-1}$  &  $M_r < -20.6$ &  16 \\
                                                     & $N=$ 5, $\Delta v = $ 1000 km s$^{-1}$ & $M_r < -20.6$ &  11 \\
                                                     & $N=$ 5, $\Delta v = $ 1000 km s$^{-1}$ &  $M_r < -20$   &  12 \\ 
                                                     & $N=$ 5, 10, 20, $\Delta v= 1000$ km s\per & $I_{AB} < 25$ & 17 \\
                                                     & $N=$ 10  & $M_V < -20$ & 18 \\
                                                     & $N=$ 10 &  $I < -24$ &  15 \\ 
                                                     & $N=$ 10, in clusters  & $M_b < -19$ & 19 \\
\hline
Galaxy number density in sphere       & $r\simeq$1 \hinv Mpc & $r < 17.77$  & 20 \\
\,\,\,\,\, of proper radius $r$       & $r=$ 8 \hinv Mpc, $\Delta v \leq 800$ km s\per & $r < 17.77$ & 21, 22, 23 \\
\hline
Number of neighbours in cylinders     & $r=$ 0.1 - 10 \hinv Mpc, $\Delta v = 1000$ km s\per & $M_{0.1r} - 5 \textrm{Log}_{10}h < -19$ &  24, 25 \\
\,\,\,\,\, with projected radius $r$  & $r=$ 0.5, 1, 2 \hinv Mpc, $\Delta v = 1000$ km s\per  & $M_r < -20$ & 26 \\
                                      & $r=$ 1 \hinv Mpc, $\Delta v$ corresponding to 8 Mpc & $r < 17.77$  &  27  \\    
                                      & $r=$ 1 - 10 \hinv Mpc, $\Delta v = 1000$ km s\per & $I_{AB} < 25$ & 17 \\
                                      & $r=$ 2 \hinv Mpc, $\Delta v = $ 1000 km s$^{-1}$ & $r < 17.77$ &  28 \\
\hline
Mass density due to nearest neighbour  & $N =$ 1 or $N$ for which $\rho$ is maximal & $M_{r, \textrm{ngb}} \gtrsim M_{r, \textrm{gal}} + 0.5 $ & 29 \\ 
\,\,\,\,\, ($\rho = 3 M_\textrm{ngb} / 4 \pi r_\textrm{ngb}^3$) & \,\,\,\,\, $\Delta v = $400, 600 km s\per & & \\
\hline
Projected galaxy number density in & \{0.5,1,2\} $< R/($\hinv Mpc $) <$ \{1,2,3\} & $M_r < -20$ & 26 \\
\,\,\,\,\, annuli                  & 1 $< R/($\hinv Mpc $)< 3$ & $r < 17.77$ & 28 \\
\hline
\hline
\textit{From simulations}  & &    \\ 
\hline
Halo mass & - & $M > 2.35 \times 10^{10}$\msun & 30 \\
\hline
Number of neighbours in spheres of radius $R$  & $R=2$ \hinv Mpc & $V_\textrm{max} > 120$ km s\per & 31 \\
\hline
Mass or density in spheres of radius $R$ & $R=1,2,4,8$ \hinv Mpc & - & 32, 33 \\
                                         & $R=5$ \hinv Mpc & - & 34, 35 \\
                                         & $R=5, 8$ \hinv Mpc & - & 36 \\
                                         & $R=7$ \hinv Mpc & - & 30 \\
                                         & $R=18, 25$ \hinv Mpc  & - & 37 \\
\hline
Matter density in spherical shells& $2 < R/($\hinv Mpc$) < 5$ & - & 38, 39, 40 \\
  & $2 < R/($\hinv Mpc$) < 7$ & - & 30 \\
  &  $R_\textrm{FOF} < R < 2$ \hinv Mpc  & - & 30 \\
  & $R_\textrm{vir} < R < 3 R_\textrm{vir}$ & - & 41 \\
\hline
Average mass density of surrounding halos & $N=$ 7 & 200  $< V_\textrm{max}/$km s\per$ < 300$ & 42 \\
\hline
Distance to nearest halo with minimum mass & - &  $M_2/M_1 > 3$ & 35 \\
\hline
\hline                                   
\end{tabular}
\end{table*}

Two main classes of observational parameters can be identified: those which measure the number of galaxies out to a given distance, and those that measure the distance out to a given $N^\textrm{th}$ neighbour. Note that the number of galaxies out to a distance $r$ is equivalent to the local number density of that same sample of galaxies smoothed on the scale $r$ with a top-hat filter. Similarly, the distance to the $N^\textrm{th}$ nearest neighbour, $r_N$, is equivalent to the local number density of galaxies smoothed on scale $r_N$. In higher density regions the $N^\textrm{th}$ nearest neighbour is, on average, closer by and the scale on which the environment is measured is therefore smaller, while the other class of methods measures the density on a fixed scale.

The environmental parameters used in simulation studies are sometimes similar to the ones used for observations, but can also be very different. Using a similar definition allows one to directly compare models and observations. However, with the full (dark matter and baryonic) density field available, simulators can also determine parameters like the total amount of mass in spheres around the galaxy in question. Such quantities might influence the evolution of a galaxy, but are difficult or impossible to obtain observationally.

It is well known that higher mass galaxies preferentially live in higher density environments. A correlation between halo mass and environmental density is therefore expected. For example, \citet{kauffmann04} and \citet{berrier10} used semi-analytic models of galaxy formation to show how their measure of environmental density (number of galaxies within \{0,5, 1, 2, 3, 6\} \hinv Mpc projected, and a redshift difference less than 1000 km s\per) correlates with halo mass. They find a good correlation with a spread of a factor of a few (for small projected cylinders) to a few tens (for larger projected cylinders). It is, however, unlikely that halo mass is the only characteristic of the environment that matters. With that in mind, \citet{fakhourima09} tried to construct an environmental parameter that does not scale with halo mass. They found that the mean over-density in a sphere of 7 Mpc, excluding the mass of the halo, gives the most mass-independent parameter of the three parameters they studied. They did not quantify the degree of correlation, but their plots indicate a weak, but non-negligible correlation with host halo mass. Observationally, this quantity cannot be determined. As far as we are aware, to date no study has found a measure of environment that is independent of halo mass.


\section{Popular environmental parameters and their relation to halo mass} \label{sec:sims}

In this section we will investigate the relation between some of the widely used environmental parameters and the mass of the host halo. For the environmental parameters discussed, we will distinguish between the `ideal case' in which the three-dimensional locations and the masses of all galaxies are known (as in simulations) (Section~\ref{sec:idealcases}), and the case in which only projected distances and velocity differences can be measured and only luminosities are available, as is the case for observations (Section~\ref{sec:projected}). First we will briefly summarise the main characteristics of the synthetic galaxy populations used.

\subsection{Simulations} 
We will compare different environmental parameters using the galaxy catalogue constructed using the semi-analytic model of \citet[][see also \citealt{croton06}]{deluciablaizot07}, run on the dark matter-only Millennium Simulation \citep{millennium}. The merger trees on which these galaxy models are built, are derived from Friends-of-Friends haloes (assuming a linking length parameter of 0.2) which are subsequently decomposed in bound subhaloes by the SubFind algorithm. The Millennium Simulation follows the evolution of the dark matter distribution using 2160$^3$ particles in a periodic volume of 500 comoving \hinv Mpc down to redshift 0. The model of \citet{deluciablaizot07} uses recipes for the evolution of the baryons inside dark matter haloes and is based on the halo merger trees constructed from the halo catalogues of the Millennium Simulation. The model predicts the galaxies' locations, physical properties such as their stellar masses and star formation histories, and observables like colours and luminosities. The model is calibrated to reproduce the redshift zero luminosity function in the $K$- and $b_J$-bands. \citet{deluciablaizot07}, \citet{delucia07}, \citet{croton06} and \citet{kitzbichlerwhite07} showed that this model reproduces many other observed properties of the galaxy population in the local Universe (e.g.\ the colour distributions, the stellar mass function and the clustering properties). We will only use the $z=0$ results. We note that the amplitude of fluctuations in the Millennium Simulation is higher than in the currently favored cosmology ($\sigma_8 = 0.9$ vs.\ 0.8). This difference may result in differences in clustering of dark matter haloes between the Millennium Simulation and the observed Universe. However, as shown by \citet{croton06}, the precursor of the galaxy formation model used here reproduced the observed clustering properties very well and the model used here does not differ much from its precursor.

We take into account all galaxies with stellar masses greater than $10^{10} $\msun. This is roughly the same lower mass limit as \citet{fakhourima09} used (they used a total mass of 1.2$\times 10^{12}$\msun). This choice is dictated by the resolution limit of the simulation. \citet{boylan-kolchin09} showed that the subhalo abundance of haloes in the Millennium Simulation is converged for subhaloes more massive than about $10^{11}$\msun, roughly independent of parent halo mass (as long as the virial mass of the parent dark matter halo is larger than $10^{12}$\msun). \citet{guo10} also investigated the subhalo abundance convergence of the Millennium Simulation and concluded that the halo and subhalo abundances are converged for $M_\textrm{vir} > 10^{12.1}$\msun. These halo masses were matched by \citet{guo10} to the stellar mass function from the seventh data release of SDSS from \citet{liwhite09}, from which they conclude that observed galaxies with stellar mass $M_* \ga 10^{10.2}$\msun\ reside in converged haloes. The number of neighbours counted in some volume depends on the lower stellar mass limit for galaxies in the sample (or, correspondingly, the flux limit of the survey), but as we will show, the scalings and correlations are usually not sensitive to this lower limit.


\subsection{The ideal case: using 3-dimensional distances and masses} \label{sec:idealcases}

We will use the simplest version of both classes of observationally determined parameters: the number of galaxies, $N_R$,  within some volume with radius $R$ and the distance to the $N^\textrm{th}$ nearest neighbour, $R_N$. Parameters derived from these numbers (such as the number density of galaxies within that volume) will obey the same conclusions. 

\begin{figure*}
   \resizebox{\textwidth}{!}{\includegraphics{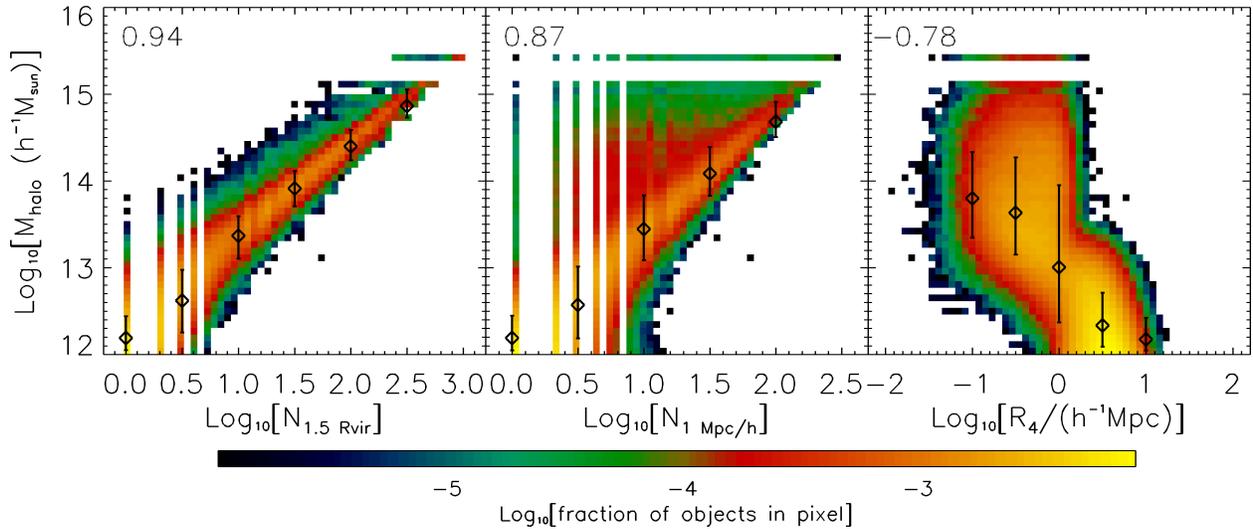}}
    \caption{The correlations between host halo virial mass and three environmental parameters: the number of galaxies within 1.5 $R_\textrm{vir}$ (left panel), the number of galaxies within 1 $h$\per Mpc (middle panel) and the distance to the fourth nearest neighbour (right panel) for all galaxies with $M_* > 10^{10}$\msun. The numbers printed in the top-left of each panel indicate the Spearman rank correlation coefficient and the diamond symbols and error bars show the binned medians and the 16th and 84th percentiles (for Gaussian distributions this would be $\pm$1$\sigma$) of the distribution. All three parameters are measures of halo mass. From left to right the correlations with halo mass decrease in strength.}{\label{fig:parametersmix}
    }     
\end{figure*}

In Fig.~\ref{fig:parametersmix} we show the correlations between host halo virial mass and three definitions of environment: the number of galaxies within 1.5 virial radii \citep[we use the definition of][throughout the paper]{bryannorman98} of the galaxies' host haloes, the number of galaxies within 1 $h^{-1}$Mpc, and the distance to the fourth nearest neighbour (left to right). All three measures of environment are strongly correlated with the mass of the host halo.

\begin{figure*}
   \resizebox{\textwidth}{!}{\includegraphics{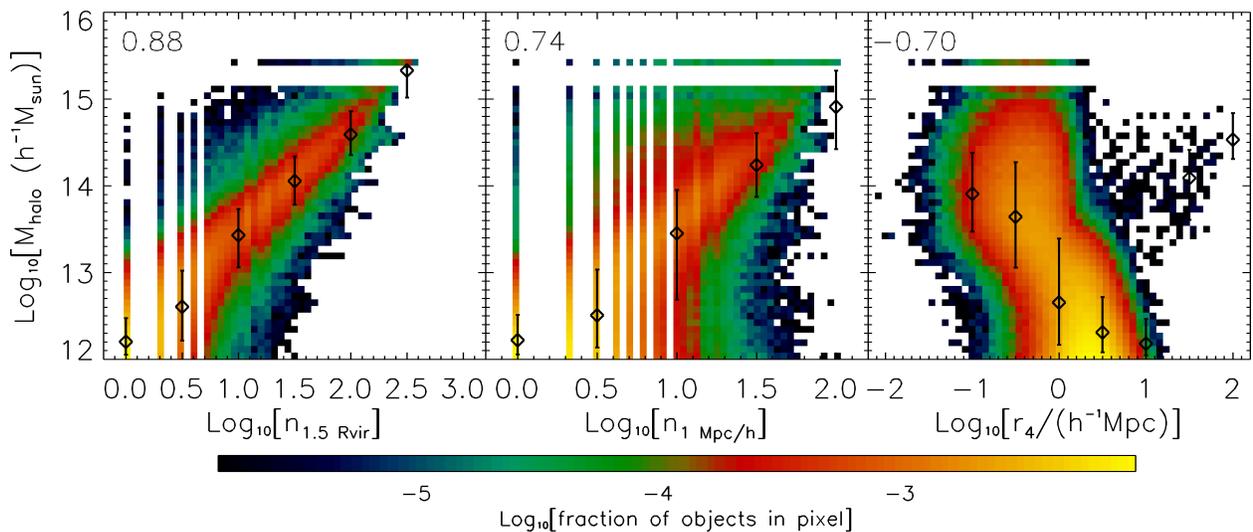}}
    \caption{As Fig.~\ref{fig:parametersmix}, but now for observable versions of the environmental parameters: We use $K$-band luminosities rather than stellar masses for the selection of galaxies in the sample and we use projected distances and a cut in redshift difference (of 1000 km s\per) rather than 3-D separations for all galaxies with $K<-23$. For the left panel, the virial radii of the host haloes still need to be known. The correlations are slightly weaker than the corresponding ones found for the ideal case (Fig.~\ref{fig:parametersmix}), mainly due to projection effects, which make galaxies populate the regions in the plots which were unoccupied in Fig.~\ref{fig:parametersmix}. The correlations are, however, still strong.}{\label{fig:2Dparametersmix}
    }     
\end{figure*}

If the distance out to which galaxies are counted is scaled to the virial radius of the parent halo that the galaxy resides in, then the correlation between halo mass and environment is very strong, as is shown in the left panel of Fig.~\ref{fig:parametersmix}. Because the region within which galaxies are counted grows with halo mass, a more or less constant fraction of satellites is counted. A fixed fraction of all satellites is a number of satellites that grows roughly linearly with halo mass, resulting in a very tight correlation. This can be understood in terms of the results found by \citet{gao04}: the fraction of the mass in subhaloes, the distribution of subhaloes and the shape of the subhalo mass function are independent of host halo mass, while the normalisation (so the total number of and total mass in subhaloes) scales (to first order) linearly with halo mass. The number of subhaloes (and thus satellite galaxies) within a radius that is fixed relative to the virial radius therefore grows roughly linearly with halo mass. This makes the parameter $N_\textrm{1.5\, Rvir}$ a very strong measure of halo mass.

A slightly weaker correlation exists between halo virial mass and the number of galaxies within a fixed physical distance, as shown in the middle panel of Fig.~\ref{fig:parametersmix} (for a distance of 1 $h^{-1}$Mpc). The upper envelope is populated by the central galaxies in the sample, while the satellites form the less tightly correlated cloud above the relation for the centrals. 

In the right panel of Fig.~\ref{fig:parametersmix} we show the correlation between host halo mass and the distance to the fourth nearest neighbour, $R_4$ (which is very often used observationally, see Table~\ref{tab:literature}). The distance $R_4$ decreases with halo mass, because more massive haloes are on average found in denser environments.

The correlation with mass is strongest for $R_4\sim 1 h^{-1}\,$Mpc, which marks the transition from the regime in which the nearest 4 galaxies are typically part of the same halo ($R_4 \ll 1 h^{-1}\,$Mpc; $M\ga 10^{13}\,$\msun) to the regime in which it resides in another halo ($R_4 \gg 1 h^{-1}\,$Mpc; $M\la 10^{13}\,$\msun).
The transition between the two regimes depends on the rank $n$: for higher ranks, the jump occurs at higher halo mass.

The three parameters displayed in Fig.~\ref{fig:parametersmix} all depend on three-dimensional distances. We will now proceed to investigate parameters that can be measured observationally.

\subsection{The realistic case: using projected distances and luminosities} \label{sec:projected}

Observationally, we have no access to the three-dimensional separations between galaxies. Instead, we measure distances projected on the sky and differences in redshift. Moreover, while luminosities are readily available, stellar mass determinations depend on SED modelling, which comes with considerable uncertainty. We will now investigate to what extent the use of observables weakens the correlations compared with the `ideal cases' discussed in Section~\ref{sec:idealcases}. As is done in many observational studies (see Table~\ref{tab:literature}), we will only make use of galaxies with redshifts that are within 1000 km s\per\ of the redshift of the galaxy for which the environment is determined. We include both the Hubble flow and peculiar velocities in our calculation of the redshifts. For reference, a velocity difference of 1000 km s$^{-1}$ corresponds to a distance of 10 $h^{-1}$Mpc if the peculiar velocity difference is zero. We will denote the parameters using the same symbols as we used for the 3-D distance variants, but with lower case letters. For example, $r_4$ denotes the projected distance to the fourth nearest neighbour (using only galaxies within the redshift difference cut). We only include galaxies with absolute $K$-band magnitude smaller than -23, which corresponds to $M_* \approx 10^{10.2}$\msun, because in our sample of resolved haloes, the galaxy luminosity function shows signs of incompleteness for fainter galaxies. This results in a slightly smaller sample than the one used before. The luminosity function of galaxies with $M_* > 10^{10}$\msun\ shows signs of incompleteness at magnitudes fainter than $K = -23$.

Fig.~\ref{fig:2Dparametersmix} shows the relation between halo mass and the parameters shown in Fig.~\ref{fig:parametersmix}, but using projected distances and luminosities rather than 3-D distances and stellar masses. For all three parameters the correlations are slightly less strong than in the ideal case. Galaxies without any neighbours within the specified distance are assigned a number density of $\log_{10}(n) = -1$.

Note that the left panel still requires knowledge of the virial radius of the host halo of the galaxy and is therefore hard to determine observationally (we left it in for completeness). The virial radius can be estimated if a group catalogue is available, like the one by \citet{yang07} who grouped galaxies using a a friends-of-friends like algorithm. The total luminosities of the groups are then ranked and matched to a ranked list of halo masses, drawn from a halo mass function sampled in a volume equal to that of the survey. This procedure results in the assignment of a host halo mass to all galaxies in the sample. However, if such a catalogue is available, then the halo mass is of course just as well known as the virial radius, so using this environmental indicator as a measure of halo mass is not very useful.

In the middle panel of Fig.~\ref{fig:2Dparametersmix} we show the halo mass as a function of the number of galaxies within a projected distance of 1 $h^{-1}$Mpc, with a redshift difference less than $\pm$1000 km s$^{-1}$ and with $K<-23$. Compared with the 3-D version (middle panel of Fig.~\ref{fig:parametersmix}), there are now more low-mass galaxies with a high number of neighbours. This is due to projection effects. We note that the correlation coefficient is still very high ($\approx$ 0.71), so we can conclude that this environmental parameter is a strong indicator of host halo mass. The horizontal scatter (scatter in environmental parameter for fixed halo mass, so this is not the scatter indicated with the error bars) at low halo masses is dominated by projection effects, while at high masses the scatter is mainly caused by satellites in the outskirts of the halo. The scatter in the environmental indicator is smallest for halo masses of about $10^{14}$ \msun. For a given $n_{\textrm{1 Mpc}/h}$ the spread in halo masses is small for low and high values of the environmental indicator (roughly 0.3 dex) and highest for $n_{\textrm{1 Mpc}/h} \sim 10$ ($\gtrsim 0.5$ dex in halo mass).

In the right panel of Fig.~\ref{fig:2Dparametersmix} we show the projected distance to the fourth nearest neighbour with $K<-23$. Because of projection effects the bi-modal behaviour visible in the right panel of Fig.~\ref{fig:parametersmix} has been smeared out. The correlation with host halo mass is therefore slightly weaker. Because of the discontinuity in the distribution, the correlation coefficient is a function of the masses (both galaxy stellar mass and host halo mass) of the objects that are taken into account.

\subsection{A multi-scale approach}

\begin{figure*}
   \resizebox{\hsize}{!}{\includegraphics{fig3.ps}}
    \caption{Halo mass as a function of the number of galaxies in annuli with inner and outer radii of 1 and 2 Mpc, respectively. The first panel shows the full sample, while the other three panels correspond to three bins in number of galaxies within 1 Mpc, showing the lower, middle and upper one eighth of the total distribution from left to right. The galaxies with $n=0$ are placed at Log$_{10}n = -1$. Although the range of halo masses in each bin is large, the typical halo mass is higher for the bins corresponding to higher values of $n_\textrm{1 Mpc/h}$. The median $n_{\textrm{1 Mpc}/h}$  for the three bins are 1, 4 and 21, respectively, and the median $\log_{10}(M_\textrm{halo}/M_\odot)$ are 12.3, 12.7 and 14.2. Dividing the sample into narrow bins of `small-scale environment' generally reduces the strength of the correlation between $n_\textrm{1-2 Mpc/h}$ and $M_{\rm halo}$, but for large values of $n_\textrm{1-2 Mpc/h}$ it actually increases it.}{\label{fig:multiscale}
    }     
\end{figure*}

\citet{wilman10} recently measured the number density of galaxies in concentric rings of 0.5, 1.0, and 2.0 $h$\per Mpc in order to investigate trends in the $u-r$ colour distribution of galaxies with environment at several distance scales (for given small-scale density, if desired). They included all galaxies from the fifth data release of SDSS with magnitude brighter than 17.77 in the $r$-band and with a mean surface brightness within the half-light radius of $\mu_r \leq 23.0$ mag arcsec$^{-2}$. The number density of galaxies was determined in rings with radii fixed in physical coordinates. In this approach neither the mass nor the distance out to which the environment is determined scales with the properties of the galaxy in question. We therefore expect that these measures of environment vary strongly with halo mass.

We find that the Spearman rank correlation coefficient for the density in annuli with halo mass is roughly 0.5, and depends on both the width and the radius of the annulus, such that smaller radii ($\la 0.5$ Mpc) have larger correlation coefficients and wider annuli mostly show weaker correlations. The power of the method of \citet{wilman10} lies in the ability to measure residual trends of galaxy properties with large-scale (annular) environment, while controlling for the environment on some smaller scale (i.e.\ the projected number density in the inner circle). The samples are constructed by taking all galaxies around which the number density of galaxies within the inner radius of the annulus fall within some bin, and are therefore comparable to vertical slices through the middle panel of Fig.~\ref{fig:2Dparametersmix}. From this figure we can see that in such a slice, a large range of halo masses is still present.

As an example, we show in Fig.~\ref{fig:multiscale} the correlation between halo mass and the number of galaxies in annuli with inner and outer radii of 1 and 2 Mpc, respectively, for three narrow bins of the number of galaxies within 1 Mpc (projected separation, within a redshift difference of 1000 km s$^{-1}$). Each bin contains 1/8 of all the galaxies, where the second panel from the left corresponds to the 1/8 of the total galaxy population with the lowest value of $n_{\textrm{1 Mpc}/h}$. Similarly, the 3rd and 4th panels show the relation between $M_{\rm halo}$ and $n_{\textrm{1-2 Mpc}/h}$ for the 1/8 of the galaxies for which $n_{\textrm{1 Mpc}/h}$ is, respectively, in the middle and the highest of the values spanned by the total population. 
The different bins in central number density (i.e.\ the different panels) favour different halo masses, as expected from Fig.~\ref{fig:2Dparametersmix}. 
Comparing panels 2--4 with the first panel, which shows the results for the full population, we see that the correlation between $M_{\rm halo}$ and $n_{\textrm{1-2 Mpc}/h}$ is much reduced for fixed, low values of $n_{\textrm{1 Mpc}/h}$. However, for high values of $n_{\textrm{1 Mpc}/h}$ (last panel) the correlation actually becomes stronger.

The trends seen in Fig.~\ref{fig:multiscale} are a typical example of the `multi-scale' approach of \citet{wilman10}. Changing the radii of the inner and outer edges of the annuli and/or the width of the bins in central galaxy number density does not affect the qualitative conclusions drawn from Fig.~\ref{fig:multiscale}. The correlation of the number of galaxies in annuli with halo mass becomes weaker if very large distances from the galaxy in question are taken (5-10 Mpc), but that merely reflects the fact that galaxies at such large distances do not have much to do with the galaxy in question.

\section{Environment as a measure of halo mass} \label{sec:distdep}
In this section we will study the strength of the correlation between several environmental indicators and halo mass in more detail. In particular, we will determine which parameter provides the best measure of halo mass.

\begin{figure*}
   \resizebox{\hsize}{!}{\includegraphics{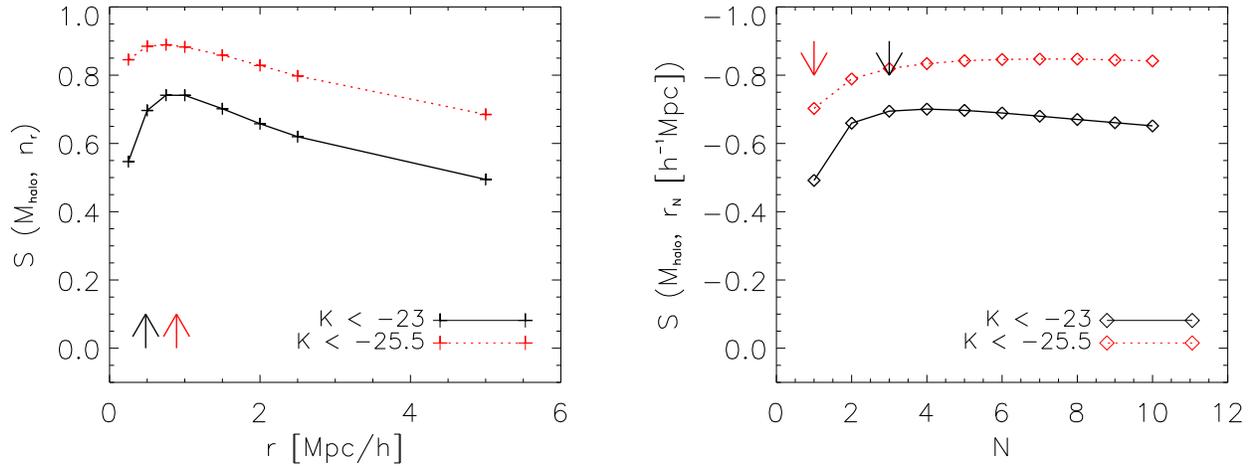}}
    \caption{The strength of the correlations between halo mass and two of the environmental indicators that can be used straightforwardly for observations for two samples of galaxies with the luminosity limits indicated in the legends. In the left panel we plot the Spearman rank correlation coefficient between halo mass and the number of galaxies within a given projected physical distance $r$ (and with a cut in redshift difference of 1000 km s$^{-1}$) as a function of $r$. The arrows show the value of the median virial radius of the haloes of all galaxies in the sample with the corresponding colours. The right panel shows the Spearman rank correlation coefficient between halo mass and the projected distance to the $N^\textrm{th}$ nearest neighbour as a function of the rank $N$. The correlation coefficient is typically negative, because more massive galaxies have their $N^\textrm{th}$ nearest neighbour closer by. The arrows indicate the median number of neighbours within the virial radius of the haloes above the indicated flux limit. The parameters $n_r$ with $r\ga 1$ Mpc and $r_N$ with $N\sim 10$ are very good measures of halo mass for a wide range of luminosities.}{\label{fig:sizedepcorr_2D}
    }     
\end{figure*}

We expect the correlation between the number of neighbours and halo mass to be strongest at some given distance. Taking the distance very small will result in strong discreteness effects, while taking the distance too large will result in a sample of galaxies that does not have much to do with the halo the galaxy resides in. 

In Fig.~\ref{fig:sizedepcorr_2D} we show, for two different environmental parameters and for two different luminosity cuts, the value of the Spearman rank correlation coefficient with halo mass, as a function of the distance related parameter used to measure the environmental density. In the left panel we show the correlation coefficient between halo mass and the environmental density indicator $n_r$ (the number of galaxies within a fixed physical distance $r$ projected on the sky and within $\Delta v = \pm 1000$ km s\per) as a function of $r$. An example of this type of parameter was shown in the middle panel of Fig.~\ref{fig:2Dparametersmix} for $r=1$ $h$\per Mpc. Fig.~\ref{fig:sizedepcorr_2D} shows that the correlation first strengthens with distance, reaches a maximum at a scale of roughly 1 \hinv Mpc, and declines slowly thereafter. The vertical arrows, which indicate the median virial radii for the haloes of all galaxies in the sample, show that the correlation between $n_{\textrm{1 Mpc}/h}$ and $M_{\rm halo}$ occurs when $r$ is slightly greater than the median virial radius. 

In the right panel of Fig.~\ref{fig:sizedepcorr_2D} we plot the Spearman rank correlation coefficient between halo mass and environment, now parametrised by $r_N$, the distance towards the $N^\textrm{th}$ nearest neighbour (as in the right panel of Fig.~\ref{fig:2Dparametersmix} for $N=4$), as a function of the rank $N$. The correlation coefficients are now mostly negative, as a higher density (corresponding to a higher halo mass) will result in a smaller distance towards the $N^\textrm{th}$ nearest neighbour. However, if we only consider galaxies with $K<-25.5$, then halo mass is an increasing function of the distance to the nearest neighbour as the neighbour needs to be outside the galaxy itself, and more massive galaxies tend to be larger. Taking more neighbours, i.e.\ higher values of $N$, gives an anti-correlation that, for high-mass galaxies, becomes stronger for larger numbers of neighbours. For lower mass galaxies ($K<-23$) the correlation between $r_N$ and $M_{\rm halo}$ is strongest for $N\approx 3-4$, but does not weaken much for larger values. The median number of neighbours within the virial radius, above the same luminosity cut is indicated with the arrows. 

We conclude that $n_r$ and $r_N$ are both good measures of host halo mass, provided that $n_r$ is measured at $r \ga r_\textrm{vir}$ and that $N$ is sufficiently large ($N\approx 10$). If the host halo mass, and thus the virial radius, is not known a priori, then it is better to take $r$ larger ($r \ga 1$ $h$\per Mpc), as the correlation rapidly weakens towards smaller distances and declines only slowly with increasing distance. Although the exact values of the correlation coefficients depend on the lower limits in halo mass and galaxy luminosity used in the study, the qualitative result that the scale for which the correlation is strongest is comparable to the median virial radius, is robust. The qualitative results are also expected to be similar at higher redshift.

\begin{figure*}
   \resizebox{\hsize}{!}{\includegraphics{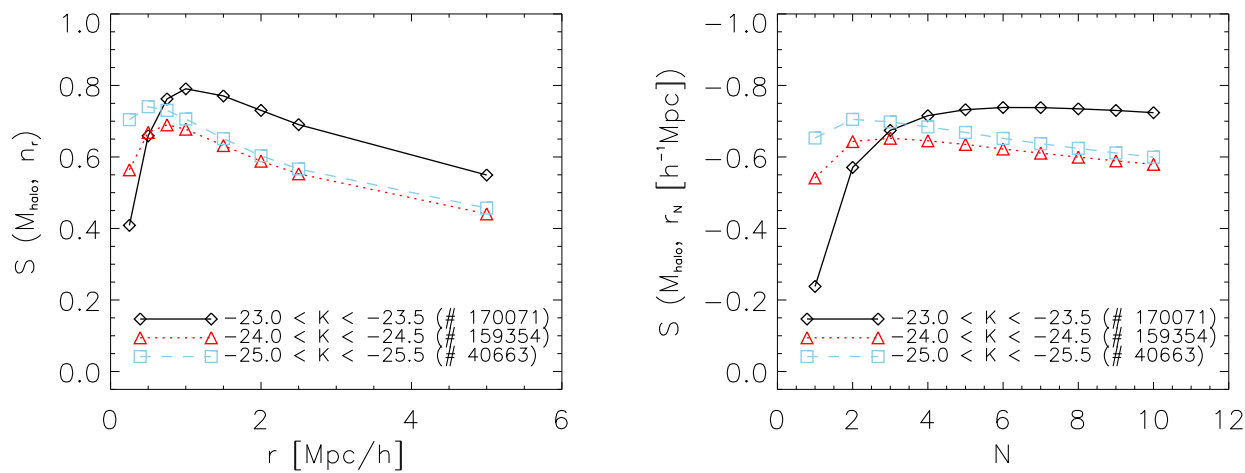}}
    \caption{The same as Fig.~\ref{fig:sizedepcorr_2D}, but now for three bins in absolute $K$-band magnitude for the galaxy under consideration (for the neighbour search all galaxies with $K<-23$ are taken into account). The numbers in between the brackets indicate the number of galaxies in the sample. The correlations between the environmental parameters and halo mass are insensitive to the $K$-band luminosity (and thus to stellar mass) except for small values of $r$ and $N$. Even for narrow bins in $K$, the environmental parameters correlate strongly with halo mass. }{\label{fig:sizedepcorr_2D_magbins}
    }     
\end{figure*}

Observational studies often contrast stellar mass and environment. A dependence on environment for fixed stellar mass is assumed to reflect nurture rather than nature. However, Fig.~\ref{fig:sizedepcorr_2D_magbins} shows that breaking up the sample in small bins of $K$-band magnitude for the galaxy under consideration (in the neighbour search we still include all galaxies with $K<-23$), which correlates well with stellar mass for central galaxies, does not reduce the strength of the correlations between environment and halo mass. Thus, effects attributed to environment, even in studies that fixed stellar mass, are mostly due to halo mass. This is important, as halo mass, at least for centrals, is considered to reflect nature rather than nurture in models of galaxy formation. 

As we will show below, using $K$-band luminosity as a proxy for (virial) mass works well. Guided by the left panel of Fig.~\ref{fig:2Dparametersmix}, one might expect that we can improve on $n_r$ as a measure of halo mass if $r$ scales with $L_K^{1/3}$. We have tried this, but the correlation between halo mass and environment does not get stronger (or it gets slightly weaker, with correlation coefficients of 0.65 -- 0.7). In the range of halo masses for which we could test it (any range between $10^{12}$ and $10^{15.5}$\msun) the correlation is stronger if a projected distance of 1 Mpc is used than if $r \propto L_K^{1/3}$ is used. Specifically, we tried $r = 1$\hinv Mpc $(L_K/L_0)^{1/3}$, with $L_0 = 10^{\{10.5, 11.0, 11.5, 12.0\}} L_\odot$. We therefore conclude that using a fixed physical projected distance is safe, and easier in practice than a distance scaling with luminosity. We thus advise to use $n_r$ with $r$ of the order of $r \gtrsim R_\textrm{vir}$, if a measure of halo mass is desired. For most observed samples of galaxies $r \sim 1$ Mpc will do, but by iteration better values can be obtained: use $r=1$ \hinv Mpc, calculate the halo virial radii from the environmental indicator (using the parametrisation given in Appendix~\ref{sec:mhalo_fits}) and then iterate if the virial radii strongly deviate from 1 Mpc.

In Appendix~\ref{sec:mhalo_fits} we provide polynomial fits for the halo mass as a function of several environmental parameters for several flux limits, which can be used to obtain halo masses from observed samples of galaxies with measured environmental indicators. Note that the fits are for $z= 0$ and may differ for higher redshift samples.


\section{Environment independent of halo mass} \label{sec:massindep}
In the previous section we have considered which environmental parameters are the best measures of halo mass. In this section we will construct new measures of environment, both for use with simulations (\S\ref{sec:massindsim}) and with observations (\S\ref{sec:massindobs}), that are highly insensitive of halo mass.

\subsection{Halo mass independent parameters for simulations}
\label{sec:massindsim}
All the environmental parameters we have looked at so far correlate with halo mass. The lower mass/luminosity limit of galaxies included as possible neighbours was set equal to the resolution limit of the simulations, or the flux limit of a survey. As we saw in the left panels of Figs.~\ref{fig:parametersmix}, \ref{fig:2Dparametersmix}, and \ref{fig:sizedepcorr_2D}, the correlation is strongest, and almost linear with halo mass, if we count neighbouring galaxies out to the virial radius of the host halo of the galaxy in question. Per unit halo mass, this galaxy number density (either projected or in a spherical region) is therefore roughly constant. This also holds for dark matter subhaloes in high-resolution simulations, as shown by \citet{gao04}.

\begin{figure}
   \resizebox{\hsize}{!}{\includegraphics{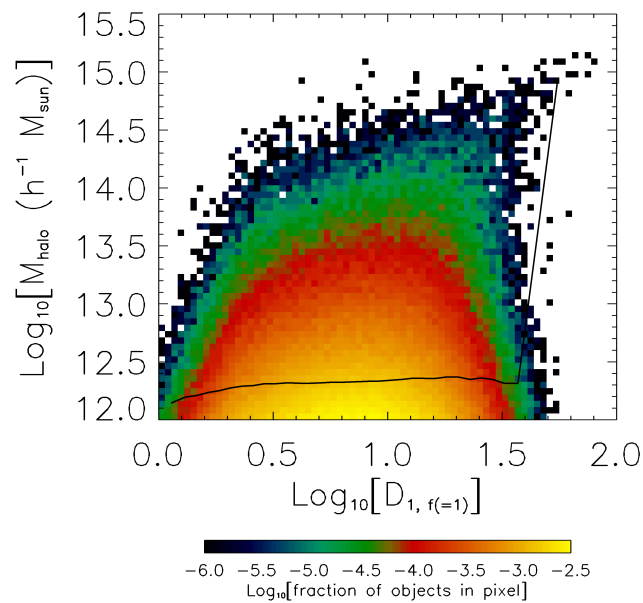}}
    \caption{Halo mass as a function of the environmental parameter $D_{1, 1}$ (eq.~\ref{eq:d_nf}). The colour scale gives the distribution for all central galaxies in the sample, while the solid line is the median halo mass in bins of $D_{1,1}$. The median relation is very flat. The correlation coefficient of this parameter with halo mass is 0.07 (for correlation coefficients as a function of rank, see Fig.~\ref{fig:sizedepcorr_massindep}). Except for the very high $D_{1,1}$ end, where the median halo mass is sufficiently high for the haloes to be on the exponential tail of the mass function, $D_{1,1}$ and halo mass are uncorrelated.}{\label{fig:massindep}
    }     
\end{figure}

In order to obtain an environmental indicator that is independent of halo mass, we expect that we will have to scale out both the mass/luminosity of the galaxy and the length scale in question. We define $D_{N, f}$ to be the three-dimensional distance to the $N^{\rm th}$ nearest neighbour with a viral mass that is at least $f$ times that of the halo under consideration, divided by the virial radius of the $N^{\rm th}$ nearest neighbour:
\begin{equation} \label{eq:d_nf}
D_{N, f} = \frac{r_{N(M_\textrm{ngb} \geq f \cdot M_\textrm{halo})}}{R_{\textrm{\scriptsize vir, ngb}}},
\end{equation}
where the subscripts `ngb' and `halo' indicate, respectively, the neighbour of the halo under consideration and the halo itself. Observe that the adjustable parameters $N$ and $f$, as well as $D_{N,f}$ itself, are all dimensionless. Because the tidal force due to the $N^{\rm th}$ nearest neighbour scales as $M_{\rm ngb}/R_N^3 \propto (R_{\rm vir,ngb}/R_N)^3$, the parameter $D_{N,f}$ scales with the tidal force to the power $-1/3$. This makes $D_{N,f}$ a natural environmental parameter with a clear physical interpretation. 

\citet{parkchoi09} found in an observational study that the physical properties of galaxies are sensitive to the distance to the nearest neighbour, normalised by the virial radius of that neighbour. They estimated virial properties of the galaxies from the luminosities. We will use a similar approach in the next section.

The colour scale of Fig.~\ref{fig:massindep} shows the distribution of haloes at $z=0$ in the $D_{1, 1} - M_\textrm{halo}$ plane. The curve shows the median $D_{1, 1}$ in bins of halo mass. Although the median $D_{1, f}$ in the sample is different for different $f$, halo mass is always independent of $D$, irrespective of the factor $f$.
The weak correlation that appears at the highest values of $D_{1, f}$ and $M_{\rm halo}$ is caused by the fact that these massive haloes are on the exponential tail of the Schechter-like halo mass function. Large-scale structure is no longer self-similar in that regime, causing a slight positive correlation between $D_{N, f}$ and halo mass. We have verified (by inverting the axes) that for masses $M \ll M_*$ (where $M_*$ is the mass at which the Schechter-like halo mass function transits from a power law into an exponential fall-off), where the mass function is a power law (and therefore scale free), the correlation is very weak. For higher masses, the exponential cut-off of the Schechter-like halo mass function imposes a mass scale. For values above roughly $f^{-1}M_{*}$, the insensitivity to mass breaks down and a weak positive correlation between halo mass and $D_{N, f}$ appears.

\begin{figure}
   \resizebox{\hsize}{!}{\includegraphics{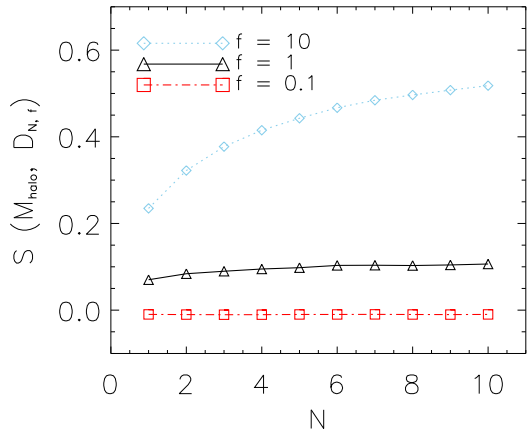}}
    \caption{The Spearman rank correlation coefficient between halo mass and the environmental indicator $D_{N, f}$ (see Eq.~\ref{eq:d_nf}) as a function of the rank $N$, for $f=\{1/10, 1, 10 \}$. Higher values for $f$ and $N$ result in stronger correlations. If haloes can be reliably identified for masses lower than the lowest mass one wants to know the environment for, then using a low value of $f$ (e.g.\ $f=0.1$) gives negligible correlations with mass. For $f=1$ the environmental parameter will be measurable for all haloes that can be identified and this value of $f$ still gives only very weak correlations with halo mass, particularly for low values of $N$. 
}{\label{fig:sizedepcorr_massindep}
    }     
\end{figure}

In Fig.~\ref{fig:sizedepcorr_massindep} we show the correlation coefficients between halo mass and $D_{N, f}$ as a function of the rank $N$ for three different values of the mass ratio of galaxies counted as neighbours and the mass of the halo under consideration, $f = \{1/10, 1, 10 \}$. For all values of $f$ the correlation between the rank $N$ and host halo mass increases with the rank\footnote{Note that the correlation does not necessarily vanish for $N\rightarrow \infty$, because neighbours must be at least $f$ times as massive as the halo under consideration. However, for very large $N$ the parameter $D_{N,f}$ no longer characterises the environment of the galaxy as the distance to the $N^{\rm th}$ nearest neighbour diverges with $N$.}.  If an environmental indicator is desired that is insensitive to halo mass, $N=1$ is therefore a good choice. The correlation is weaker for lower values of $f$. For $f<1$ the environmental indicator cannot be determined for the full resolved sample of haloes (as halo masses need to be at least $M>f^{-1} M_\textrm{res}$, with $M_\textrm{res}$ the resolution limit, in order to resolve all possible neighbours). We therefore advise to take $f=1$, as then the parameter can be defined for all galaxies in the sample and it still gives only a very weak correlation with halo mass. If in a sample of haloes some of the studied properties demand a much more stringent resolution limit (e.g. if detailed halo profiles need to be fitted), and if haloes of much lower mass are resolved in terms of their virial mass and position, then one should use values of $f<1$, e.g. 0.1, for which the correlation between halo mass and environment becomes vanishingly small.

If, in the definition of $D_{N,f}$, we replace the virial radius of the neighbour by the virial radius of the halo under consideration  (thereby losing the connection to the tidal force due to the neighbour), the correlation between halo mass and environment becomes even slightly weaker (e.g. a Spearman rank correlation coefficient of 0.04 instead of 0.07 between halo mass and $D_{1,1}$). As using the virial radius of the neighbour gives a more intuitive external environmental parameter, we prefer to use the virial radius of the neighbour. 

We conclude that the parameter $D_{N,f}$, with $N=1$ and $f\leq 1$, results in an intuitive environmental parameter that is very insensitive to halo mass. We do note, however, that in order to calculate this halo mass independent environmental indicator, one needs a measure of the virial mass of the host halo. From simulations these can easily be obtained. For observed samples of galaxies virial masses can be estimated using the environmental indicators that do correlate strongly with halo mass, as described in the previous section and detailed in Appendix~\ref{sec:mhalo_fits}. In the next section we will present an environmental indicator that can be obtained directly from observations and that is also insensitive to halo mass.

\subsection{Halo mass independent parameters for observed samples of galaxies}
\label{sec:massindobs}
\begin{figure*}
   \resizebox{\hsize}{!}{
   \includegraphics{fig8a.ps} 
   \includegraphics{fig8b.ps} 
   \includegraphics{fig8c.ps}
   }
    \caption{The host halo masses of galaxies versus the dimensionless environmental parameter $d_{1,0}$ (see Eq.~\ref{eq:d_Nm}), which is easy to measure from observations. In the left panel we show all galaxies in the sample, for which the Spearman rank correlation coefficient is -0.28. The middle and right panels show the same relation for central and satellite galaxies, respectively. For these sub-samples the Spearman rank correlation coefficients are 0.09 and -0.35, respectively. The correlation between $M_{\rm halo}$ and $d_{1,0}$ is weak and nearly vanishes for centrals.}{\label{fig:massindep_obs}
    }     
\end{figure*}

The environmental parameter $D_{N,f}$ defined in the previous section depends on the masses of the haloes, which are generally not readily available for observed samples of galaxies. Halo masses can e.g.\ be estimated by abundance matching \cite[e.g.][]{kravtsov04, valeostriker04, conroy06, conroywechsler09}, where a sorted list of galaxy luminosities is matched to a sorted list of halo masses (from either an analytic halo mass function or a simulation), with or without scatter in the one-to-one relation. However, this technique does not work for satellite galaxies and it is generally not known which galaxies are centrals and which are satellites. If a group catalogue is available, then the results can be improved by summing the luminosities of all galaxies in a group and matching the sorted list of group luminosities to the sorted list of halo masses \citet[e.g.][]{yang03, vandenbosch03, yang07}. Finally, as we have shown in Section~\ref{sec:distdep}, it is possible to use common environmental parameters to estimate halo mass. We provide detailed instructions for doing so in the Appendix.

Since it is easier to work with environmental parameters that do not require knowledge of halo masses and because $D_{N,f}$ also required knowledge of 3-dimensional distances, we set out to formulate an environmental parameter that can be easily determined observationally and that is as insensitive of halo mass as possible. We let the definition of $D_{N, f}$ guide us. We know that we have to scale the minimum masses/luminosities of the galaxies that are taken into consideration in the search for neighbours to be a fixed fraction of the mass/luminosity of the galaxy under consideration and that we have to scale the distance to the neighbours to some typical length scale associated with the neighbour.

We use an observable, the $K$-band luminosity $L_K$, instead of mass. Luminosity is easier to measure and does not require the modeling of the spectral energy distribution of the galaxy. We use the $K$-band because in the very red optical bands and in the near-IR the correlation between luminosity and stellar mass is strongest \citep[aside from the uncertainties arising from the treatment of thermally pulsing asymptotic giant branch stars, see e.g.][]{maraston05, tonini10}. Instead of 3-dimensional distances we will use projected distances, considering only neighbours with redshifts that are within 1000 km s$^{-1}$ of the galaxy for which we are measuring the environment. Instead of dividing by the virial radius of the neighbour, we divide by $L_K^3$, as $R_{\rm vir}\propto M^{1/3}$ and we expect that $L_K$ will scale roughly as $M_{\rm halo}$, at least for central galaxies. Note that this last assumption is known to break down \citep[e.g.][]{kravtsov04, cooray06, conroywechsler09}, but we find that using more complicated functional forms to fit the relation between $R_{\rm vir}$ and $(L_K)$ does not significantly change the correlation between environment and halo mass. 

Our environmental indicator $d_{N,m}$ then becomes
\begin{equation} \label{eq:d_Nm}
d_{N, m} = \frac{ r_{\textrm{N}(K \leq K_\textrm{gal}-m)}}{0.58 h^{-1} \textrm{Mpc}} \cdot \biggr( \frac{L_{K, \textrm{ngb}}}{1.4 \times 10^{11} L_\odot} \biggr)^{-1/3}
\end{equation}
where the subscript `ngb' again denotes the neighbour of the galaxy in question, $m$ is the difference in magnitudes (corresponding to a ratio in luminosity/mass, a positive $m$ implies that the neighbours must be brighter) between the galaxy in question and the galaxies counted as possible neighbours, and $K$ is the absolute $K$-band magnitude. Although the coefficients do not affect the correlation with halo mass, we have divided the projected distance by 
$R_\textrm{vir,13} = 0.58 h^{-1}\,{\rm Mpc} (L_K / 1.4\times 10^{11}\,L_\odot)^{-1/3}$ rather than just by $L_K^{1/3}$, where $R_\textrm{vir,13} = 0.58 h^{-1}$Mpc is the projected virial radius of the `reference mass' of $10^{13}$\msun, to ensure that the actual values of $d_{N,m}$ retain some intuitive, physical meaning.
If $R_\textrm{vir,13} (L_K/1.4 \times 10^{11} L_\odot)^{1/3}$ were the virial radius, then the external environmental indicator $d_{N, m}$ could be described as the projected distance to the $N^\textrm{th}$ nearest neighbour that is at least $m$ magnitudes brighter than the galaxy we are measuring the environment of, normalised to the neighbour's projected virial radius.

\begin{figure}
   \resizebox{\hsize}{!}{\includegraphics{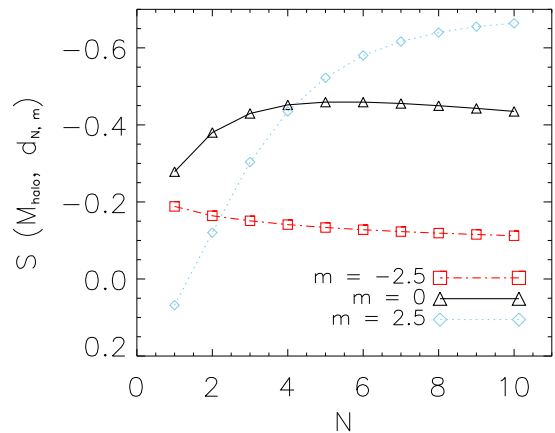}}
    \caption{The Spearman rank correlation coefficient between halo mass and $d_{N, m}$ as a function of the rank $N$, for $m=\{-2.5, 0, 2.5\}$ magnitudes (i.e.\ neighbours with $K$-band luminosities that are at least 0.1, 1.0, and 10 times as large as that of the galaxy under consideration). If an environmental indicator that is insensitive to halo mass is desired, then $d_{1,2.5}$ and $d_{N,0}$ for any $N\la 10$ are all good choices.
}{\label{fig:sizedepcorr_obsmassindep}
    }     
\end{figure}

The colour scale in the left panel of Fig.~\ref{fig:massindep_obs} shows the distribution of galaxies in the $M_\textrm{halo} - d_{1, 0}$ plane. We include all galaxies in the catalogue with $K < -23$. 
The Spearman rank correlation coefficient between halo mass and environment is -0.28, which indicates a weak anti-correlation. 

The middle and right panels of Fig.~\ref{fig:massindep_obs} show the distribution of central galaxies and satellites, respectively. For these sub-samples the Spearman rank correlation coefficient between $d_{1,0}$ and halo mass are 0.09 and -0.35, respectively. Thus, the weak anti-correlation seen in the left panel is mostly due to the inclusion of satellites. More massive haloes contain more satellites above a fixed magnitude limit. Hence, for satellites in more massive haloes, the distance to the nearest brighter galaxy, which can be another satellite or the central galaxy of the same halo, is typically smaller relative to the virial radius of the halo. 

The parameter shown in Fig.~\ref{fig:massindep_obs} requires the difference in redshift between the neighbour and the galaxy under consideration to be smaller than 1000 km s$^{-1}$. Without this cut in redshift difference the correlations become stronger. Taking into account only galaxies within a redshift window is important, but the result is not very sensitive to the precise velocity cut.

The dependence of the correlation between host halo mass and $d_{N, m}$ on the rank $N$ is shown in Fig.~\ref{fig:sizedepcorr_obsmassindep}, for three different values of $m$. We have chosen to show $m=\{-2.5, 0, 2.5 \}$ magnitudes, because a magnitude difference of 2.5 corresponds to a luminosity ratio of 10, similar to the mass ratio of 10 that we used earlier. If neighbours that are less luminous than the galaxy under consideration are allowed to count as neighbours (i.e.\ $m$ negative), then the sample for which the $d_{N,m}$ can be determined is smaller than the total sample of galaxies (because all possible, lower mass neighbours need to be resolved) and the typical haloes the galaxies are in are more massive. For $m=-2.5$, i.e.\ neighbours that are at least 0.1 times as bright, we find a weak anti-correlation that slowly decreases in strength with increasing $N$, from $S\approx -0.2$ for $N=1$ to $S\approx -0.1$ for $N=10$. For $m=0$ the anti-correlation is weakest for $N=1$, the case shown in Fig.~\ref{fig:massindep_obs}. For $m=2.5$, i.e.\ neighbours that are at least 10 times more luminous than the galaxy under consideration, we find a very small, positive correlation for $N=1$ ($S\approx 0.07$), which turns into a anti-correlation for $N=2$ that quickly increases in strength with increasing $N$. Hence, if an environmental indicator that is insensitive to halo mass is desired, then $d_{1,2.5}$ and $d_{N,0}$ for any $N\la 10$ are all good choices.

\section{Conclusions} \label{sec:conclusions}
The properties of observed galaxies and dark matter haloes in
simulations depend on their environment. The term ``environment'' has,
however, been used to describe a wide variety of measures that may or
may not correlate with each other. Useful measures of environment
include, for example, the distance to the $N^{\rm th}$ nearest
neighbour, the number density of objects within some distance, or, for
the case of galaxies, the mass of the host dark matter halo. In this paper we carried out a detailed investigation of several environmental parameters which are popular in the (observational) literature, focusing in particular on their relationship with halo mass. 

We measured the environmental indicators from the synthetic galaxy catalogues produced using the semi-analytic model by \citet{deluciablaizot07}, built on the Millennium Simulation \citep{millennium}. This model reproduces the number density and clustering properties of observed galaxies in the low-redshift Universe. All our results hold for galaxies with a stellar mass in excess of $10^{10}$\msun\ in haloes above $10^{12}$ \msun\ at $z \sim 0$. At lower masses and higher redshifts the results are expected to be qualitatively similar.

We showed that it is of crucial importance to realise that the degree to which environmental parameters measure host dark matter halo mass, is determined by (1) whether the distance out to which the environment is measured scales with some typical length scale (e.g.\ the virial radius of the halo hosting the neighbour) and (2) whether or not the minimum mass/luminosity that the neighbours are required to have is fixed in absolute terms or relative to the mass/luminosity of the galaxy in question. Specifically, we found that
\begin{enumerate}
\item All frequently used environmental indicators (i.e.\ some function of the distance to the $N^\textrm{th}$ nearest neighbour or the number of galaxies within some given distance, either using three-dimensional distances or using projected distances for all galaxies within some radial velocity difference) correlate strongly with halo mass. This remains true if only galaxies within some narrow range of ($K$-band) luminosities are considered.
\item For the number of galaxies within a distance $r$, $n_r$, the correlation with halo mass is strongest if we set $r$ equal to 1.5--2 virial radii. The virial radius is generally difficult to measure from observations and knowing it would remove the need to measure halo mass, but the correlation with halo mass is nearly as strong for galaxy counts within $\sim1$ Mpc. 
\item  The strength of the anti-correlation between the distance to the $N^\textrm{th}$ nearest neighbour, $r_N$, and halo mass is nearly constant for $N\ge 2$ and only slightly weaker for $N=1$. The relation between halo mass and $r_N$ is slightly weaker than that between halo mass and $n_r$ if $r$ is taken to be similar to the virial radius.
\item Both $n_r$ and $r_N$ correlate more strongly with halo mass if the neighbours are required to be more luminous or massive.
\end{enumerate}

We have shown that it is possible to construct environmental parameters that are highly insensitive to halo mass by using only dimensionless quantities. For the case of dark matter haloes in numerical simulations this can for example be achieved by measuring $D_{N,f}$, the distance to the $N^{\rm th}$ nearest halo, that is at least $f$ times as massive as the halo under consideration, divided by the virial radius of that neighbour. Dividing by the virial radius of the halo hosting the galaxy itself yields even slightly weaker correlations with halo mass, but dividing by the virial radius of the neighbour gives the indicator an intuitive interpretation: the tidal force due to the neighbour scales as $1/D_{N,f}^3$. The correlation with halo mass becomes weaker if the minimum mass required for neighbours is lower (i.e.\ for lower $f$). These environmental parameters are, however, only insensitive to halo mass for haloes that are not on the exponential tail of the mass function. 
 
In the case of observations, we usually only know a position on the sky, some rough indication of the distance along the line of sight, and the flux or luminosity in some waveband. We showed that analogous environmental measures that are highly insensitive to halo mass can also be constructed using only the $K$-band luminosities, projected distances on the sky, and a maximum radial velocity difference for neighbours. Specifically, the parameter $d_{N,m}$, defined as the projected distance to the $N^{\rm th}$ nearest galaxy that is at least $m$ magnitudes brighter and that is within a radial velocity difference of 1000 km s$^{-1}$, divided by the $K$-band luminosity of that neighbour to the power one third, correlates only very weakly with host halo mass for suitable choices of $N$ and $m$ (e.g.\ $m=2.5$ and $N=1$ or $m=0$ and any $N\le 10$). The correlation vanishes nearly completely for samples that only contain central galaxies.

In summary, when measuring environments for (virtual) observations, we advise to make use of both a halo mass independent measure and a measure that is highly sensitive to halo mass. For purely theoretical studies the halo mass is already known and we therefore advise to use an environmental parameter that is insensitive to halo mass. The following parameters are good choices: 
\begin{itemize}
\item \emph{Insensitive to halo mass; for simulations:} The distance to the nearest (main) halo that is at least $f$ times more massive than the halo in question, divided by the virial radius of that neighbour. The choice $f=1$ works well, but if resolution permits it, smaller values yield even weaker correlations with halo mass.
Dividing instead by the virial radius of the halo itself gives a slightly weaker correlation with halo mass, at the expense of losing the intuitive definition in which the environment relates to the tidal force due to the neighbour.
\item \emph{Insensitive to halo mass; for observations:} The parameter $d_{1,0}$, as given by Eq.~\ref{eq:d_Nm}. The correlation with halo mass is weaker if satellites are excluded. 
\item \emph{Sensitive to halo mass; for observations:} The number of brighter galaxies within a projected distance of $\sim 1$ \hinv Mpc, within a redshift window corresponding to $\Delta v \lesssim 1000$ km s\per ($n_\textrm{1 Mpc/h}$). Even better would be to subsequently iterate the following two steps until the procedure converges: (i) check what the corresponding halo masses are using the relations between $n_r$ and halo mass given in Appendix~\ref{sec:mhalo_fits}; (ii) adapt the maximum projected distance to 1.5 times the typical virial radius of the haloes in the sample.
\end{itemize}

Many studies have measured galaxy properties as a function of both stellar mass and environment. We have shown that the environmental indicators used by most authors are effectively measures of halo mass, even for fixed $K$-band luminosity, which is a proxy for stellar mass. While halo mass is a perfectly valid measure of environment, and may be particularly relevant for satellites, we note that because stellar mass is also expected to correlate strongly with halo mass, these studies may not have separated ``internal'' and ``external'' influences as well as one might naively think. The work presented here will enable future observational and theoretical studies to disentangle the effects of halo mass (internal environment) from those of the external environment. This may eventually tell us whether halo mass is the only important driver of the physics governing galaxy evolution.

\section*{Acknowledgements}
The authors kindly thank the anonymous referee for constructive comments, Claudio Dalla Vecchia for assistance in the early stages of this study and for a careful reading of the manuscript. We thank Nadieh Bremer for spotting a minor bug in an earlier version of our code. It is also a pleasure to thank Jarle Brinchmann and Scott Trager for constructive comments on an early version of this paper. The Millennium Simulation databases used in this paper and the web application providing online access to them were constructed as part of the activities of the German Astrophysical Virtual Observatory. We thank John Helly for his support in our efforts to access these databases. This work was supported by an NWO VIDI grant and by the Marie Curie Initial Training Network CosmoComp (PITN-GA-2009-238356).


\bibliographystyle{mn2e}
\bibliography{marcelsbib}

\appendix

\section{Obtaining the halo mass from environmental parameters} \label{sec:mhalo_fits}

In this Appendix we describe the tables we provide in the electronic edition of the paper and on the website http://environment.marcelhaas.com. These tables provide fitting functions for the halo mass as a function of different environmental indicators.

We will use environmental parameters that can be obtained directly from observations and those that need an iterative scheme outlined below, in order to estimate the virial radius of the host haloes of the galaxies in question. We provide the parameters corresponding to third order polynomial fits for the halo mass as a function of the environmental indicators. We fit a function of the form
\begin{equation} \label{eq:fit}
\log_{10} [M_\textrm{halo} (h^{-1} M_\odot)] = A + B P + C P^2 + D P^3
\end{equation}
Where $P$ indicates the logarithm (base 10) of the environmental parameter in question. For all indicators we fit to the median halo masses in bins separated by $\Delta P = 0.25$.

The fitted values for the four polynomial coefficients ($A, B, C, D$) are given in the online tables (see sample table~\ref{tab:fits_obs}) for nine different environmental parameters ($r_1$, $r_4$, $r_{10}$, $n_{\textrm{0.5 Mpc}/h}$, $n_{\textrm{1 Mpc}/h}$, $n_{\textrm{2 Mpc}/h}$, , $n_{\textrm{1 Rvir}}$, $n_{\textrm{1.5 Rvir}}$ and $n_{\textrm{2 Rvir}}$).

Because the distribution in halo masses at fixed environment is neither Gaussian nor symmetric, we provide similar fits to the 16$^\textrm{th}$ and 84$^\textrm{th}$ percentiles of the data set (these percentiles would correspond to $\pm 1 \sigma$ for Gaussian distributions), such that at all values for the environment the median and 16$^\textrm{th}$ and 84$^\textrm{th}$ percentiles of the distribution of halo masses can be found.

We provide tables for K-band magnitude limited samples with galaxies brighter than $K = \{-23, -23.5, -24, -24.5, -25, -25.5\}$ and tables for stellar mass limited samples with $\log_{10}[M_* (h^{-1} M_\textrm{sun})] \geq \{10, 10.2, 10.4, 10.6, 10.8, 11\}$. In the neighbour search we include all galaxies above the given flux or stellar mass limit, with a maximum redshift difference of 1000 km/s between the neighbour and the galaxy we determine the environmental parameter for. 

For all parameters and (stellar mass or flux limited) samples we provide fits for the halo mass including all galaxies, or in small bins of stellar mass or $K-$band magnitude. Note that the obtained halo masses may become unreliable below $10^{12} M_\textrm{sun}$, the convergence limit of the simulation, and that the fits were derived for $z=0$. For higher redshifts and lower halo masses, a similar analysis would need to be carried out.

\subsection{Iterating towards the best halo mass estimator}
As we have shown in Section~\ref{sec:idealcases}, the strongest correlation between halo mass and environment is obtained when galaxies are counted within a distance that scales with the virial radius of the halo hosting the galaxy under consideration. In order to do so, an estimate of the host halo mass is required, which can be obtained from the relations described earlier in this Appendix. Using
\begin{equation}
R_\textrm{vir} = 0.27 \, \textrm{\hinv Mpc} \, \Biggr(\frac{M_\textrm{halo}}{10^{12} \textrm{\msun}} \Biggr)^{1/3}  \frac1{1+z},
\end{equation}
which is the relation that was used in this work to obtain virial radii, an estimate for the virial radius can then be obtained. Here $z$ is the redshift, which is zero throughout this work.

The best estimate of the halo mass can then be found by measuring the projected number of neighbours within a given multiple of the virial radius (with the same cut in radial velocity difference), as shown in Section~\ref{sec:projected}. In the tables we provide the same third order polynomial fits, but for the relation between halo mass and $n_\textrm{1 Rvir}$, $n_\textrm{1.5 Rvir}$ an $n_\textrm{2 Rvir}$, as well as the corresponding (higher) Spearman rank correlation coefficients.

This procedure for obtaining a better estimate for the halo mass can then be iterated towards a reliable estimate for the halo mass, including the spread in halo masses at fixed environment (note that this spread is very small for high-mass haloes if the neighbours are counted within a multiple of the virial radius of about one.)

Finally, we caution that these halo masses are measured from the Millennium Simulation, which uses the WMAP first-year results for the cosmology, which has (among other differences) too large an amplitude of fluctuations ($\sigma_8=0.9$). This means that for a given galaxy luminosity, the haloes will be slightly too massive. How this affects the relations between environment and halo mass is not clear.

\label{lastpage}

\begin{landscape}
\begin{table*}
\caption{A sample table from the complete tables provided in the online edition and on our website (http://environment.marcelhaas.com). This table contains the fits for halo mass as a function of environment for all galaxies brighter than $K = -23$ and is shown here for two of the environmental parameters. The first two columns give the upper and lower flux limits for which the fits are made ($K=-999$ indicates no upper flux limit was used). Note that in the neighbour search all galaxies brighter than $K=-23$ are used (within a redshift difference of 1000 km s\per). Columns 3 - 6 are the coefficients $A - D$ of Eq.~\ref{eq:fit} for the median relation (subscript `med'), and columns 7 - 10 and 11 - 14 are the coefficients $A - D$ of Eq.~\ref{eq:fit} for the 16$^\textrm{th}$ and 84$^\textrm{th}$ percentiles (subscripts `16' and `84'), respectively. The last column gives the Spearman rank correlation coefficient between host halo mass and the environmental parameter. Note that all lines for a given environmental parameter, except for the first and last, can be used to determine the host halo mass from the environment for fixed $K$ band luminosity (and for fixed stellar mass in the corresponding tables).}             
\label{tab:fits_obs}      
\centering                          
\begin{tabular}{l c c c c c c c c c c c c c c}        
\hline                
\hline
 $K_\textrm{max}$ & $K_\textrm{min}$ & $A_\textrm{med}$ & $B_\textrm{med}$ & $C_\textrm{med}$ & $D_\textrm{med}$ & $A_{16}$ & $B_{16}$ & $C_{16}$ & $D_{16}$ & $A_{84}$ & $B_{84}$ & $C_{84}$ & $D_{84}$ & $S(M_\textrm{halo}, P)$ \\
\hline 
$P = \log_{10} [r_1 (h^{-1}$Mpc$)]$ & & & & & & & & & & & & & & \\
\hline
 -23.0 &  -999.0 &  12.4 &   -0.51  &   0.64  &   0.39  &  12.02  &  -0.21  &   0.55  &   0.29  &  12.96  &    -1.0  &    0.6  &  0.5  &  -0.49  \\
 -23.0 &   -23.5 &  12.4 &   -1.05  &   0.54  &   0.55  &  11.98  &  -0.41  &   0.55  &   0.37  &  12.96  &    -1.5  &    0.5  &  0.6  &  -0.24  \\
 -23.5 &   -24.0 &  12.4 &   -0.55  &   0.57  &   0.38  &  12.01  &  -0.20  &   0.55  &   0.29  &  12.97  &    -1.1  &    0.5  &  0.5  &  -0.48  \\
 -24.0 &   -24.5 &  12.2 &   -0.65  &   0.90  &   0.57  &  11.89  &  -0.24  &   0.87  &   0.45  &  12.73  &    -1.3  &    0.9  &  0.7  &  -0.54  \\
 -24.5 &   -25.0 &  12.2 &   -0.66  &   0.79  &   0.48  &  11.89  &  -0.31  &   0.77  &   0.41  &  12.70  &    -1.1  &    0.7  &  0.6  &  -0.58  \\
 -25.0 &   -25.5 &  12.3 &   -0.67  &   0.51  &   0.30  &  12.09  &  -0.33  &   0.34  &   0.16  &  12.73  &    -0.9  &    0.5  &  0.4  &  -0.65  \\
 -25.5 &  -999.0 &  12.4 &   -0.88  &   0.75  &   0.42  &  12.17  &  -0.46  &   0.61  &   0.24  &  12.69  &    -1.2  &    0.7  &  0.5  &  -0.70  \\
\hline
$P = \log_{10} [r_4  (h^{-1}$Mpc$)]$  & & & & & & & & & & & & & & \\
\hline
 -23.0 &  -999.0 &  12.8  &  -0.87  &   0.51  &   0.30  &  12.37  &  -0.89   &  0.49  &   0.33  &  13.40  &    -0.9  &    0.4  &  0.3  &  -0.70  \\
 -23.0 &   -23.5 &  12.8  &  -1.02  &   0.56  &   0.42  &  12.34  &  -1.01   &  0.55  &   0.41  &  13.38  &    -1.1  &    0.5  &  0.4  &  -0.72  \\
 -23.5 &   -24.0 &  12.8  &  -0.85  &   0.47  &   0.29  &  12.38  &  -0.92   &  0.47  &   0.37  &  13.44  &    -0.9  &    0.4  &  0.3  &  -0.70  \\
 -24.0 &   -24.5 &  12.8  &  -0.79  &   0.57  &   0.29  &  12.29  &  -0.88   &  0.58  &   0.36  &  13.37  &    -0.9  &    0.5  &  0.3  &  -0.65  \\
 -24.5 &   -25.0 &  12.6  &  -1.16  &   0.73  &   0.49  &  12.29  &  -0.97   &  0.60  &   0.43  &  13.21  &    -1.1  &    0.6  &  0.5  &  -0.62  \\
 -25.0 &   -25.5 &  12.7  &  -1.08  &   0.50  &   0.35  &  12.33  &  -0.90   &  0.58  &   0.31  &  13.18  &    -1.3  &    0.4  &  0.4  &  -0.68  \\
 -25.5 &  -999.0 &  12.9  &  -1.43  &   0.31  &   0.47  &  12.58  &  -1.26   &  0.47  &   0.48  &  13.36  &    -1.4  &    0.1  &  0.3  &  -0.83  \\
\hline
...   & & & & & & & & & & & & & & \\
...   & & & & & & & & & & & & & & \\
                 
\end{tabular}
\end{table*}
\end{landscape}

\end{document}